\documentclass{article}
\usepackage{amsfonts}
\usepackage{amsmath}

\newtheorem{theorem}{Theorem}
\newtheorem{remark}[theorem]{Remark}
\makeatletter
\@addtoreset{equation}{section}
\makeatother

\begin{document}

\title{A discrete Schr\"{o}dinger spectral problem\\
and associated evolution equations}
\author{M. Boiti, M. Bruschi${}^\dag$, F. Pempinelli and B. Prinari \\
Dipartimento di Fisica dell'Universit\`{a} and Sezione INFN,\\
Lecce, Italy\\
${}^\dag$Dipartimento di Fisica dell'Universit\`a di Roma ``La Sapienza",\\
Roma, Italy}
\date{11 June, 2002}
\maketitle

\begin{abstract}
A recently proposed discrete version of the Schr\"{o}dinger spectral problem
is considered. The whole hierarchy of differential-difference nonlinear
evolution equations associated to this spectral problem is derived. It is
shown that a discrete version of the KdV, sine-Gordon and Liouville
equations are included and that the so called `inverse' class in the
hierarchy is local. The whole class of related Darboux and B\"{a}cklund
transformations is also exhibited.
\end{abstract}

\section{Introduction}

In recent years there has been a growing interest in the field of discrete
dynamical systems, i.e. systems that can be described by ordinary
and{$\backslash$}or partial, generally nonlinear, difference equations.

Such systems arise and play an important role in a very large number of
contexts and have an extensive range of applications: mathematical physics,
chaos, fractals, disordered systems, biology, optics, economics, statistical
physics, numerical analysis, discrete geometry, cellular automata, quantum
field theory and so on\footnote{
The related literature is so large that it is impossible to quote even if in
a very partial way. The interested reader can find a lot of relevant
references in J. of Phys. A, \textbf{34} n. 48 (2001), Special issue:\emph{\
Symmetries and integrability of difference equations (SIDE IV)}}.

Even if powerful analytic tools were developed in the last decades to deal
with difference equations, due to the nonlinearity of the systems of
interest, no technique of solution is available for most of such dynamical
systems. Thus it is clear the relevance and the interest of any new
integrable one.

In this paper we introduce a whole class of interesting nonlinear
differential-difference evolution equations which correspond to isospectral
deformations of the new Schr\"{o}dinger discrete spectral problem introduced
by \cite{Shabat} and investigated in \cite{KdVdiscrete} and which,
therefore, are integrable via the inverse scattering method. We show that in
this class are included the new discrete integrable version of the
celebrated KdV found in \cite{KdVdiscrete}, the discrete sine-Gordon and
Liouville equations and a whole hierarchy of local equations, which as far
as we know is new. We give also the explicit recurrence operators to
construct the Darboux Transformations and the B\"{a}cklund Transformations
for the whole class. Of course the B\"{a}cklund transformations are
interesting `per se' as difference-difference dynamical systems and moreover
they can be suitable to construct special solitonic solutions for the
considered differential-difference equations.

\section{The spectral problem}

Let us consider the spectral problem
\begin{equation}
L(n)\psi (n;\lambda )=\lambda \psi (n;\lambda )  \label{sp}
\end{equation}
with
\begin{equation}
L(n)=E^{2}+q(n)E^{1}  \label{L}
\end{equation}
where $n$ is a discrete variable\ ($n\in
\mathbb{Z}
$), $\lambda \in \mathbb{C}$ is the spectral parameter and $E^{k}$ is the
`shift' operator defined by
\begin{equation}
E^{k}\phi (n)=\phi (n+k),\quad \quad k=0,\pm 1,\pm 2,...  \label{E}
\end{equation}
This spectral equation is the discretized version of the Schr\"{o}dinger
equation recently obtained by Shabat \cite{Shabat} iterating Darboux
transformations of the continuous Schr\"{o}dinger equation and its direct
and inverse problems were first introduced and studied in \cite{KdVdiscrete}.

In the following the `potential' $q$ and consequently the eigenfunction $
\psi $ will be considered depending as well on the continuous time variable $
t$, while the spectral parameter is considered time independent, so that we
can apply the inverse scattering method and consider the associated
diffferential-difference nonlinear equations.

In the following we shall often use the shorthand notation
\begin{equation}
\phi (n)=\phi ,\quad \quad \phi (n+k)=\phi _{k},\quad \quad \quad k=\pm
1,\pm 2,...  \label{shn}
\end{equation}

We will use also the first order difference operators

\begin{align}
\Delta & =E^{1}-E^{0}  \label{inv+-} \\
\widetilde{\Delta }& =E^{1}+E^{0}  \label{invtilde}
\end{align}
and their inverses $\Delta ^{-1},\widetilde{\Delta}^{-1}$. Useful formulas for
such difference operators and for difference equations are reported in
Appendix A.

\section{Isospectral hierarchy}

\label{hierarchy}We are looking for nonlinear discrete evolution equations
associated with the isospectral deformations of the discrete Schr\"{o}dinger
operator $L$ introduced in (\ref{L}). They can be obtained from the Lax
representation
\begin{equation}
\dot{L}=\left[ L,M\right]  \label{lax}
\end{equation}
where $M(n,t,E^{k})$ is an opportune shift operator and dot denotes time
differentiation. We have
\begin{equation}
\dot{q}(n,t)E^{1}=V(n,t)E^{1}
\end{equation}
where $V(n,t)=V(q,q_{k})\ $is a function depending on $q$ and its shifted
values.

To construct the hierarchy of isospectral nonlinear discrete evolution
equations we can use a sort of dressing procedure. Precisely, we look for
the \emph{recursion operators} $\mathcal{M}$ and $\mathcal{L}$ which allow
to construct new admissible operators $M^{\prime }$ and $V^{\prime }$ in
terms of known $M$ and $V$. Following the technique introduced in \cite{db}
and \cite{db1}, we start with the ansatz
\begin{equation}
M^{\prime }=LM+AE^{0}+BE^{1}  \label{m1}
\end{equation}
where $E^{0}$ is the identity operator and $A$ and $B$ are functions,
depending on the $q$ and $q_{k},$ to be determined in such a way that
\begin{equation}
V^{\prime }E^{1}=[L,M^{\prime }]
\end{equation}
or explicitly
\begin{equation}
V^{\prime }E^{1}=q\left( A_{1}-A\right) E^{1}+\left(
A_{2}-A+qB_{1}-q_{1}B+qV_{1}\right) E^{2}+\left( B_{2}-B+V_{2}\right) E^{3}.
\label{v2}
\end{equation}

Now imposing for compatibility the vanishing of the terms in $E^{2}$ and $
E^{3}$ and setting, for convenience,
\begin{equation}
C=A_{1}-A  \label{x0}
\end{equation}
we get the two conditions
\begin{align}
& B_{2}-B=-V_{2}  \label{x1} \\
& C_{1}+C=q_{1}B-qB_{1}-qV_{1}  \label{x2}
\end{align}
The general solution of the second order difference equation (\ref{x1})
reads
\begin{equation}
B=\overline{b}+\widehat{b}\left( -1\right) ^{n}+\sum\limits_{k=0}^{\infty
}V_{2k+2}  \label{x3}
\end{equation}
where $\overline{b},\widehat{b}$ are two arbitrary constants.

Taking into account (\ref{x3}) the general solution of the first order
difference equation (\ref{x2}) reads
\begin{align}
C& =\widetilde{C}-q\sum\limits_{k=0}^{\infty
}V_{2k+1}-2\sum\limits_{k=1}^{\infty }\left( -1\right)
^{k}q_{k}\sum\limits_{j=0}^{\infty }V_{k+2j+1}  \label{x4a} \\
\widetilde{C}& =\check{c}\left( -1\right) ^{n}-\overline{b}\left(
q+2\sum\limits_{k=1}^{\infty }\left( -1\right) ^{k}q_{k}\right) +\widehat{b}
\left( -1\right) ^{n}\left( q+2\sum\limits_{k=1}^{\infty }q_{k}\right)
\label{x4b}
\end{align}
where $\check{c}$ is again an arbitrary constant.

Now, from (\ref{v2}), (\ref{x0}), (\ref{x4a}) and (\ref{x4b}) we get
\begin{equation}
V^{\prime }=\widetilde{V}+\mathcal{L}V  \label{r1}
\end{equation}
where the \emph{recursion operator} $\mathcal{L}$ is given by
\begin{equation}
\mathcal{L}V=-q\left( q\sum\limits_{k=0}^{\infty
}V_{2k+1}+2\sum\limits_{k=1}^{\infty }\left( -1\right)
^{k}q_{k}\sum\limits_{j=0}^{\infty }V_{k+2j+1}\right)  \label{rr}
\end{equation}
while
\begin{equation}
\widetilde{V}=q\left[ \check{c}\left( -1\right) ^{n}-\overline{b}\left(
q+2\sum\limits_{k=1}^{\infty }\left( -1\right) ^{k}q_{k}\right) +\widehat{b}
\left( -1\right) ^{n}\left( q+2\sum\limits_{k=1}^{\infty }q_{k}\right)
\right] .  \label{r0}
\end{equation}

Note that starting from the trivial operators $M=0,V=0$ we get from (\ref{r1})
the non trivial starting point $\widetilde{V}$. Indeed, taking into
account the fact that $\overline{b},\widehat{b},\breve{c}$ are \emph{arbitrary}
constants, we have three different independent starting points
\begin{align}
\breve{V}& =\left( -1\right) ^{n}q  \label{sp1} \\
\overline{V}& =q\left( q+2\sum\limits_{k=1}^{\infty }\left( -1\right)
^{k}q_{k}\right)  \label{sp2} \\
\widehat{V}& =\left( -1\right) ^{n}q\left( q+2\sum\limits_{k=1}^{\infty
}q_{k}\right)  \label{sp3}
\end{align}
so that the class of isospectral nonlinear discrete evolution equations
associated with the spectral operator (\ref{sp}), (\ref{L}) is a
superposition of three hierarchies and reads
\begin{equation}
\dot{q}=\alpha \left( \mathcal{L}\right) \breve{V}+\beta \left( \mathcal{L}
\right) \overline{V}+\gamma \left( \mathcal{L}\right) \widehat{V}  \label{c1}
\end{equation}
where $\alpha ,\beta ,\gamma $ are entire functions of their argument. The
possibility of using negative powers of the recursion operator will be
considered below.

It is easily seen that the recursion operator $\mathcal{L}$ in (\ref{rr})
can be written, by using the difference operators (\ref{inv+-}), (\ref{invtilde})
and their inverses (see Appendix A), in the more compact and
elegant form
\begin{equation}
\mathcal{L}=-q\Delta \widetilde{\Delta }_{{}}^{-1}qE^{1}\Delta ^{-1}
\widetilde{\Delta }^{-1}.  \label{CalligraphicL}
\end{equation}

The explicit form of the recurrence operator $\mathcal{M}$ for $M$ ($
M^{\prime }=\mathcal{M}M\iff V^{\prime }=\mathcal{L}V)$ can be easily
obtained from the above formulas (see (\ref{m1}), (\ref{x0}), (\ref{x3}),
(\ref{x4a}), (\ref{x4b})).

\begin{remark}
Note that all the $M$ operators generated by the recursion (\ref{m1})
contain only positive powers of the shift operator $E$.
\end{remark}

\begin{remark}
Due to the non-local character of the recursion operator $\mathcal{L}$
(see (\ref{rr})) and taking into account the starting points (\ref{sp1}),
(\ref{sp2}), (\ref{sp3}), all the equations in the class (\ref{c1}) are apparently non
local (except the rather trivial $\dot{q}=\check{V}=\left( -1\right) ^{n}q$).
However it is easily seen that $\dot{q}=\overline{V}=q\left(
q+2\sum\limits_{k=1}^{\infty }\left( -1\right) ^{k}q_{k}\right) $ and $\dot{
q}=\widehat{V}=\left( -1\right) ^{n}q\left( q+2\sum\limits_{k=1}^{\infty
}q_{k}\right) $ imply respectively
\begin{align}
\dot{q}q_{1}+q\dot{q}_{1}& =qq_{1}\left( q-q_{1}\right)  \label{qqpoint1} \\
\dot{q}q_{1}+q\dot{q}_{1}& =(-1)^{n}qq_{1}\left( q_{1}+q\right)
\label{qqpoint2}
\end{align}
which are local although not explicit. The first equation was derived and
studied in ref. \cite{KdVdiscrete}.
\end{remark}

\section{The `inverse' class}

As noted in the previous section, the equations in the class (\ref{c1})
correspond to positive shifts in the Lax operator $M$. In this section we
investigate the class of isospectral nonlinear discrete evolution equations
which corresponds to operators $M$ constructed in terms of negative powers
of the shift operator $E$.

These equations can be obtained using the inverse of the recursion operator $
\mathcal{L}$, namely
\begin{equation}
\mathcal{L}^{-1}=-E^{-1}\Delta \widetilde{\Delta }\frac{1}{q}\Delta ^{-1}
\widetilde{\Delta }\frac{1}{q}.  \label{rinv}
\end{equation}

Thus, given a valid couple $M$ and $V$, one can obtain a new valid one $
M^{\prime }=\mathcal{M}^{-1}M$, $V^{\prime }=\mathcal{L}^{-1}V.$

Starting from $V=0$ and taking into account that an arbitrary constant $c$\
is in the $\emph{ker}$ of $\Delta $ we have
\begin{equation}
\widetilde{V}=\mathcal{L}^{-1}0=-2~c\left( \frac{1}{q_{1}}-\frac{1}{q_{-1}}
\right)  \label{eqinv11}
\end{equation}
and the whole `inverse' class of isospectral nonlinear discrete evolution
equations reads
\begin{equation}
\dot{q}=\eta \left( \mathcal{L}^{-1}\right) \widetilde{V}
\end{equation}
where $\eta $ is an arbitrary entire function of its argument.

However, in order to study the properties of this class of equations, it is
more convenient to construct it explicitly through the ansatz
\begin{equation}
M^{\left( N\right) }=\sum\limits_{k=1}^{N}A^{(k)}E^{k-N-1}.  \label{mn1}
\end{equation}
Inserting the above expression in the Lax equation (\ref{lax}) it is easily
seen that the functions $A^{(k)}$ $(k=1,2,...N)$ are determined recursively
by
\begin{equation}
qA_{1}^{(s+1)}-q_{s-N}A_{{}}^{(s+1)}=A^{(s)}-A_{2}^{(s)},\quad
A^{(0)}=0,\quad \quad s=0,1,...,N-1.  \label{ra}
\end{equation}
The corresponding nonlinear discrete evolution equation reads
\begin{equation}
\dot{q}=A_{2}^{(N)}-A^{(N)}.  \label{eqi1}
\end{equation}
The general solution of (\ref{ra}) is, according to (\ref{AXBXF}) and (\ref{solutionAXBXF})
\begin{equation}
A_{{}}^{(s+1)}=\frac{\overline{a}^{(s+1)}+\sum\limits_{k=0}^{\infty }\left(
A_{k+2}^{(s)}-A_{k}^{(s)}\right) \displaystyle\frac{1}{q_{k}}
\prod\limits_{j=~1}^{N-s}q_{-j+k+1}}{\prod\limits_{j=~1}^{N-s}q_{-j}}
\label{As+1}
\end{equation}
where the constants $\overline{a}^{(s+1)}$ arise from the solution of the
homogeneous part of the equation and can be chosen equal to zero except $
\overline{a}^{(1)}=1$ that gives a convenient starting point for the
iteration.

In spite of the infinite series in (\ref{As+1}) the $A^{(s+1)}$ are local
for any $N$. In the Appendix B we solve explicitly the recursion relation
proving by induction that
\begin{equation}
A_{{}}^{(s+1)}=\frac{(-1)^{s}}{\prod\limits_{m=~1}^{N-s}q_{-m}}
\sum_{k_{1}=-N}^{-1}\;\sum_{k_{2}=-N+1}^{k_{1}+1}\dots
\sum_{k_{j}=-N+j-1}^{k_{j-1}+1}\dots
\sum_{k_{s}=-N+s-1}^{k_{s-1}+1}Q_{k_{1}}\dots Q_{k_{s}},
\end{equation}
where
\begin{equation}
Q_{k}=\frac{1}{q_{k}q_{k+1}}.
\end{equation}
For the reader's convenience, we write here explicitly the first three
equations in the hierarchy.

\begin{enumerate}
\item For $N=1$ we get
\begin{equation}
\dot{q}=\left( \frac{1}{q_{1}}-\frac{1}{q_{-1}}\right)  \label{eqinv1}
\end{equation}
which was implicitly considered in \cite{KdVdiscrete}. Note that setting
\begin{equation}
q=\frac{1}{u}  \label{pos1}
\end{equation}
we get
\begin{equation}
\dot{u}=-u^{2}\left( u_{1}-u_{-1}\right) .
\end{equation}

\item For $N=2$ we get
\begin{equation}
\dot{q}=-\left( \frac{1}{q_{1}^{2}}\left( \frac{1}{q_{2}}+\frac{1}{q}\right)
-\frac{1}{q_{-1}^{2}}\left( \frac{1}{q_{-2}}+\frac{1}{q}\right) \right)
\label{eqinv2}
\end{equation}
which, using the position (\ref{pos1}), becomes
\begin{equation}
\dot{u}=u^{2}\left( u_{1}^{2}\left( u_{2}+u\right) -u_{-1}^{2}\left(
u_{-2}+u\right) \right)
\end{equation}

\item For $N=3$ we get
\begin{align}
\dot{q}& =\left( \alpha _{2}-\alpha \right)   \label{eqinv3a} \\
\alpha & =\frac{1}{qq_{-1}^{2}}\left( \frac{1}{q_{1}q}+\frac{1}{qq_{-1}}+
\frac{1}{q_{-1}q_{-2}}\right) +\frac{1}{q_{-1}^{2}q_{-2}}\left( \frac{1}{
qq_{-1}}+\frac{1}{q_{-1}q_{-2}}+\frac{1}{q_{-2}q_{-3}}\right) .
\label{eqinv3b}
\end{align}
With the position (\ref{pos1}) we get the polynomial equation
\begin{align}
\dot{u}& =-u^{2}\left( \beta _{2}-\beta \right)  \\
\beta & =uu_{-1}^{2}\left( u_{1}u+uu_{-1}+u_{-1}u_{-2}\right)
+u_{-1}^{2}u_{-2}\left( uu_{-1}+u_{-1}u_{-2}+u_{-2}u_{-3}\right) .
\end{align}
\end{enumerate}

\section{Discrete KdV, sine-Gordon and Liouville equations}

Notice that the first equation (\ref{eqinv1}) in the `inverse' class of the
equations related to the discrete Schr\"{o}dinger operator (\ref{L}) can be
rewritten as
\begin{equation}
q_{1}\dot{q}+\dot{q}_{1}q=\frac{q}{q_{2}}-\frac{q_{1}}{q_{-1}}
\end{equation}
and, then, by taking a linear combination with coefficients $c$ and $d$ of
this equation with the equation (\ref{qqpoint1}) of the direct class we get
\begin{equation}
q_{1}\dot{q}+\dot{q}_{1}q=c\left( \frac{q}{q_{2}}-\frac{q_{1}}{q_{-1}}
\right) +d(q_{1}-q)q_{1}q
\end{equation}
which is the equation studied in \cite{KdVdiscrete} and that for $c=2d$
reduces to a discrete version of the KdV equation.

It was already shown in \cite{sinediscrete} that the discrete sine-Gordon
and Liouville equations are included in the hierarchy of integrable
equations related to the discrete Schr\"{o}dinger spectral operator (\ref{L}).
They can be recovered by using the special dressing method described
above. In fact if, instead of starting in the iteration procedure from $V=0$,
$M=0$, we start from
\begin{equation}
M=\partial _{t},\quad V=-\dot{q}  \label{w1}
\end{equation}
we get
\begin{equation}
\dot{q}=V^{\prime }=-\mathcal{L}\dot{q},  \label{ee}
\end{equation}
which is a tricky equation since in order to extract from it an evolution
equation it must be solved with respect to $\dot{q}$. In order to do this it
is convenient to put
\begin{equation}
\dot{q}=-\gamma \Delta \widetilde{\Delta }e^{-2\varphi }  \label{qpointphi}
\end{equation}
with $\gamma $ an arbitrary constant and $\varphi =\varphi (n,t)$ a new
function and
\begin{equation}
q=-e^{2\varphi _{1}}\widetilde{\Delta }P  \label{qB}
\end{equation}
with $P=P(n,t)$ to be determined. Inserting (\ref{qpointphi}) and (\ref{qB})
into (\ref{ee}) we have
\begin{equation}
P_{1}^{2}-P^{2}=e^{-2(\varphi _{1}+\varphi _{2})}-e^{-2(\varphi +\varphi
_{1})}.
\end{equation}
With an opportune choice of the constant of integration and of the sign of $
P $, we get
\begin{equation}
P=e^{-(\varphi +\varphi _{1})}.
\end{equation}
Therefore,
\begin{equation}
q=-e^{\varphi _{1}-\varphi }-e^{\varphi _{1}-\varphi _{2}}  \label{qphi}
\end{equation}
and the differential-difference equation in $\varphi $ can be obtained by
imposing the compatibility between (\ref{qpointphi}) and (\ref{qphi}). We get
\begin{equation}
(\dot{\varphi}_{1}-\dot{\varphi})e^{\varphi _{1}-\varphi }-(\dot{\varphi}_{2}
-\dot{\varphi}_{1})e^{\varphi _{1}-\varphi _{2}}=\gamma e^{-2\varphi
_{2}}-\gamma e^{-2\varphi }.
\end{equation}
Inserting in it
\begin{equation}
\dot{\varphi}_{1}-\dot{\varphi}=-\gamma e^{-(\varphi _{1}+\varphi )}+\Phi
\end{equation}
with $\Phi =\Phi (n)$ to be determined we derive
\begin{equation}
\Phi _{1}e^{-(\varphi _{1}+\varphi _{2})}=\Phi e^{-(\varphi +\varphi _{1})}
\end{equation}
that can be integrated furnishing the evolution equation
\begin{equation}
\dot{\varphi}_{1}-\dot{\varphi}=-\gamma e^{-(\varphi _{1}+\varphi )}+\gamma
^{\prime }e^{(\varphi _{1}+\varphi )}
\end{equation}
which, up to a trivial change of function, for $\gamma ^{\prime }=\gamma $
is the discrete sine-Gordon and for $\gamma ^{\prime }=0$ the Liouville
equation.

Also the auxiliary spectral problem introduced in \cite{sinediscrete} for
fixing the time evolution of $\varphi $ can be recovered by using our
dressing method. In fact we have
\begin{equation}
\dot{\psi}=-M^{\prime }\psi  \label{w5}
\end{equation}
which, since
\begin{equation}
M^{\prime }=LM+AE^{0}+BE^{1},  \label{w2}
\end{equation}
can be rewritten as
\begin{equation}
\dot{\psi}=-[L,M]\psi +\lambda M\psi -A\psi -B\psi _{1}.
\end{equation}
Recalling that $M=\partial _{t}$ and denoting $\lambda =-1-k^{2}$, we have
\begin{equation}
k^{2}\dot{\psi}=A\psi +(B-\dot{q})\psi _{1}.  \label{psiev}
\end{equation}
Using the formulas obtained for $A$ and $B$ in section \ref{hierarchy} and
choosing equal to zero the constants of integration we get
\begin{equation}
A=-\widetilde{\Delta }_{{}}^{-1}qE^{1}\Delta ^{-1}\widetilde{\Delta }^{-1}V
\label{w3}
\end{equation}
and
\begin{equation}
B=-E^{2}\Delta ^{-1}\widetilde{\Delta }^{-1}V.  \label{w4}
\end{equation}
Inserting the formulas for $q$ and $\dot{q}$ obtained in (\ref{qphi}) and
(\ref{qpointphi}) we recover the auxiliary spectral problem derived in \cite{sinediscrete}
\begin{equation}
k^{2}\dot{\psi}=\gamma e^{-\varphi _{1}-\varphi }\psi -\gamma e^{-2\varphi
}\psi _{1}.
\end{equation}
The second auxiliary problem introduced in \cite{sinediscrete} is obtained
by closing the compatibility conditions with the discrete Schr\"{o}dinger
spectral problem (\ref{sp}).

\section{Darboux and B\"{a}cklund transformations}

Let us consider besides (\ref{sp}) a second spectral problem with a
different `potential', namely let
\begin{equation}
\widetilde{L}\widetilde{\psi }=\lambda \widetilde{\psi }
\end{equation}
with
\begin{equation}
\widetilde{L}(n,t)=E^{2}+\widetilde{q}(n,t)E^{1}.
\end{equation}
Let us introduce a Darboux transformation relating $\psi $ and $\widetilde{
\psi }$
\begin{equation}
\widetilde{\psi }=\mathcal{D}\psi
\end{equation}
where $\mathcal{D}$ is an opportune shift operator depending on $q$ and $
\widetilde{q}$. This implies a relation between $q$ and $\widetilde{q}$,
called B\"{a}cklund transformation, that can be expressed in the following
operatorial form (see e.g. \cite{db}, \cite{db1})
\begin{equation}
\widetilde{L}\mathcal{D-D}L=WE^{1}=0
\end{equation}
where the scalar operator $W$ depends on $q$ and $\widetilde{q}$ and their
shifted values up to some order.

Now, following a technique introduced in \cite{db1}, we look for the
\emph{recursion operators} $\Gamma $ and $\Omega $ which allow to construct a
valid couple of $\mathcal{D}^{\prime }$ and $W^{\prime }$ from the supposed
known $\mathcal{D}$, $W$, that is such that $\mathcal{D}^{\prime }=\Gamma
\mathcal{D}$, $W^{\prime }=\Omega W$.

Consider the following \emph{ansatz}
\begin{equation}
\mathcal{D}^{\prime }=\widetilde{L}\mathcal{D}+FE^{0}+GE^{1}
\end{equation}
where $F~$and $G$ are functions to be determined requiring that
\begin{equation}
W^{\prime }E^{1}=\widetilde{L}\mathcal{D}^{\prime }\mathcal{-D}^{\prime }L
\end{equation}
for some $W^{\prime }$. We have
\begin{equation}
W^{\prime }E^{1}=\left( \widetilde{q}F_{1}-qF\right) E^{1}+\left( F_{2}-F+
\widetilde{q}G_{1}-q_{1}G+\widetilde{q}W_{1}\right) E^{2}+\left(
G_{2}-G+W_{2}\right) E^{3}  \label{q1}
\end{equation}
and imposing the vanishing of the terms in $E^{2}$ and $E^{3}$ we get the
following conditions
\begin{align}
G_{2}-G& =-W_{2}  \label{bca} \\
F_{2}-F& =q_{1}G-\widetilde{q}G_{1}-\widetilde{q}W_{1}.  \label{bcb}
\end{align}

The general solution of the second order difference equation (\ref{bca})
reads
\begin{equation}
G=g+\overline{g}\left( -1\right) ^{n}-E^{2}\Delta ^{-1}\widetilde{\Delta }
^{-1}W  \label{G}
\end{equation}
where $g$, $\overline{g}$ are two arbitrary constants.

The general solution of the second order difference equation (\ref{bcb})
reads
\begin{equation}
F=f+\overline{f}\left( -1\right) ^{n}+\Delta ^{-1}\widetilde{\Delta }^{-1}
\left( q_{1}G-\widetilde{q}G_{1}-\widetilde{q}W_{1}\right)
\end{equation}
where $f$, $\overline{f}$ are two arbitrary constants.

Therefore, inserting (\ref{G}) we have
\begin{equation}
F=\widetilde{F}+\overline{F}  \label{F}
\end{equation}
where
\begin{equation}
\widetilde{F}=f+\overline{f}\left( -1\right) ^{n}-gE^{1}\Delta ^{-1}
\widetilde{\Delta }^{-1}(\widetilde{q}_{-1}-q)-\overline{g}
(-1)^{n}E^{1}\Delta ^{-1}\widetilde{\Delta }^{-1}\left( \widetilde{q}_{-1}
+q\right)  \label{Ftilde}
\end{equation}
and
\begin{equation}
\overline{F}=E^{1}\Delta ^{-1}\widetilde{\Delta }^{-1}\left\{ -\widetilde{q}_{-1}
+(\widetilde{q}_{-1}E^{1}-q)E^{1}\Delta ^{-1}\widetilde{\Delta }
^{-1}\right\} W.  \label{Fbar}
\end{equation}

From (\ref{q1}), (\ref{F}), (\ref{Ftilde}) and (\ref{Fbar}) we get
\begin{equation}
W^{\prime }=\widetilde{W}+\Lambda W  \label{WLambda}
\end{equation}
where
\begin{align}
\widetilde{W}& =\left( \widetilde{q}\widetilde{F}_{1}-q\widetilde{F}\right)
\\
\Lambda & =\left( \widetilde{q}E^{1}-q\right) E^{1}\Delta ^{-1}\widetilde{
\Delta }^{-1}\left\{ -\widetilde{q}_{-1}+(\widetilde{q}_{-1}E^{1}-q)E^{1}
\Delta ^{-1}\widetilde{\Delta }^{-1}\right\} .
\end{align}

Starting from the trivial operators $\mathcal{D}=0,W=0$ we get from (\ref{WLambda})
a non trivial B\"{a}cklund transformation $\widetilde{W}$. Indeed,
taking into account the arbitrariness of the constants in (\ref{Ftilde})
we have four different independent starting points so that the class of
B\"{a}cklund transformations reads
\begin{equation}
\sum\limits_{k=1}^{4}\alpha _{k}\left( \Lambda \right) W^{(k)}=0
\end{equation}
where the $\alpha _{k}\left( \Lambda \right) $ are \emph{arbitrary} entire
functions of their argument and the `elementary' B\"{a}cklund
transformations $W^{(k)}$ are given by
\begin{align}
W^{(1)}& =\left( \widetilde{q}-q\right) \\
W^{(2)}& =(-1)^{n}\left( \widetilde{q}+q\right) \\
W^{(3)}& =\left( \tilde{q}E^{1}-q\right) E^{1}\Delta ^{-1}\widetilde{\Delta }^{-1}
\left( \widetilde{q}_{-1}-q\right) \\
W^{(4)}& =\left( -1\right) ^{n}\left( \widetilde{q}E^{1}+q\right)
E^{1}\Delta ^{-1}\widetilde{\Delta }^{-1}(\widetilde{q}_{-1}+q).
\end{align}

The explicit form of the recurrence operator $\Gamma $ for $\mathcal{D}$ can
be easily obtained from the above formulas.

\begin{remark}
In the limit $\widetilde{q}\rightarrow q$, as expected, the operator $
\Lambda $ becomes the operator $\mathcal{L}$.

In fact let us note that $\Lambda $ can be rewritten as
\begin{equation*}
\Lambda =\left( q-\widetilde{q}E^{1}\right) \left\{ \sum_{k=0}^{\infty
}E^{2k}(q_{1}-\widetilde{q}E^{1})\sum_{j=1}^{\infty
}E^{2j}-\sum_{k=0}^{\infty }E^{2k}\widetilde{q}E^{1}\right\} .
\end{equation*}
For $\widetilde{q}=q$ we get, recalling (\ref{Delta2})
\begin{equation*}
\Lambda _{\widetilde{q}=q}=\left( q-qE^{1}\right) \sum_{k=0}^{\infty
}(-1)^{k}E^{k}qE^{1}\Delta ^{-1}\widetilde{\Delta }^{-1}
\end{equation*}
and, thanks to (\ref{Delta3Inverse}) we have
\begin{equation*}
\Lambda _{\widetilde{q}=q}=\mathcal{L}.
\end{equation*}
\end{remark}

\section{Concluding remarks}

To end this paper we want to outline a number of possible extensions and
generalizations.

First of all we conjecture that the whole hierarchy here introduced is
endowed with a double Hamiltonian structure. Finding such structure should
allow us to exhibit an infinite number of commuting conservation laws for
the whole hierarchy.

Moreover, the technique we used to derive our results (recurrence operators,
Lax pairs and Darboux transformations) can be easily extended to recover
differential-difference nonlinear evolution equations related with non
isospectral deformations of the spectral problem, getting typically
equations with $n$-dependent coefficient (see \cite{br1}).

The B\"{a}cklund transformations and the discrete-discrete evolution
equations that can be associated to this new spectral problem
(see \cite{KdVdiscrete,sinediscrete}) deserve further investigation as integrable
nonlinear iterated maps and moreover offer a good starting point for the
introduction of new `integrable' cellular automata (see e.g. \cite{brs}).

Finally, all these results could be generalized to the non-abelian case
considering a matrix discrete Schr\"{o}dinger operator and matrix
differential-difference evolution equations (see e.g. \cite{lv1}). \appendix

\section{Appendix}

For the convenience of readers not so familiar with difference equations and
operators we give here some useful formulas.

Let us define the first order difference\emph{\ }operators
\begin{align}
\Delta & =E^{1}-E^{0} \\
\widetilde{\Delta }& =E^{1}+E^{0}
\end{align}
where $E^{0}$ is the identity operator. The general solutions of the
difference equations
\begin{eqnarray}
\Delta X &=&F,\quad \quad X=X(n),\;F=F(n) \\
\widetilde{\Delta }Y &=&G\quad \quad Y=Y(n),\;G=G(n)
\end{eqnarray}
can be written as
\begin{eqnarray}
X &=&c+\Delta _{{}}^{-1}F \\
Y &=&c(-1)^{n}+\widetilde{\Delta }_{{}}^{-1}G
\end{eqnarray}
where $c$ is an arbitrary constant, i.e. not depending on $n$, and the
inverses of $\Delta $ and $\widetilde{\Delta }$ are chosen as follows
\begin{align}
\Delta _{{}}^{-1}& =-\sum\limits_{k=0}^{\infty }E^{k}  \label{Delta2Inverse}
\\
\widetilde{\Delta }_{{}}^{-1}& =\sum_{k=0}^{\infty }(-1)^{k}E^{k}.
\label{Delta3Inverse}
\end{align}
We can also construct higher order operators using these fundamental ones
as the second order difference operator
\begin{equation}
\Delta \widetilde{\Delta }=E^{2}-E^{0}  \label{prop}
\end{equation}
and its inverse
\begin{equation}
\Delta ^{-1}\widetilde{\Delta }^{-1}=-\sum_{k=0}^{\infty }E^{2k}.
\label{Delta2}
\end{equation}
Thus the general solution of the second order difference equation
\begin{equation}
\Delta \widetilde{\Delta }X=F
\end{equation}
can be written as follows
\begin{equation}
X=c+\overline{c}(-1)^{n}+\Delta ^{-1}\widetilde{\Delta }^{-1}F  \label{X2-X1}
\end{equation}
where $c$ and $\overline{c}$ are arbitrary constants.

Finally, let us consider the general homogeneous first order difference
equation
\begin{equation}
AX_{1}-BX=0,\quad \quad X=X(n),\;A=A(n),\;B=B(n).
\end{equation}
Its general solution can be written as follows
\begin{equation*}
X=c\prod_{k=0}^{+\infty }\frac{A_{k}}{B_{k}}
\end{equation*}
where $c$ is again an arbitrary constant.

Then for the non homogenous first order difference equation
\begin{equation}
AX_{1}-BX=F,\quad \quad F=F(n)  \label{AXBXF}
\end{equation}
we search a solution of the form
\begin{equation}
X=\prod_{k=0}^{+\infty }\frac{A_{k}}{B_{k}}G
\end{equation}
with $G=G(n)$ to be determined. We get
\begin{equation}
G_{1}-G=\frac{1}{B}\prod_{k=0}^{+\infty }\frac{B_{k}}{A_{k}}F
\end{equation}
and therefore
\begin{equation}
X=c\prod_{k=0}^{+\infty }\frac{A_{k}}{B_{k}}-\sum_{j=0}^{+\infty }\frac{1}{
A_{j}}\prod_{k=0}^{j}\frac{A_{k}}{B_{k}}F_{j}.  \label{solutionAXBXF}
\end{equation}

\section{Appendix}

We choose $\overline{a}^{(1)}=1$ and all other $\overline{a}^{(s+1)}=0$. It
is convenient to rewrite the recursion relation for the $A^{(s+1)}$ in terms
of
\begin{equation}
B^{(s)}=A^{(s)}\prod_{j=1}^{N-s+1}q_{-j}.  \label{B}
\end{equation}
We have
\begin{equation}
B_{1}^{(s+1)}-B^{(s+1)}=-\frac{B_{2}^{(s)}}{qq_{1}}+\frac{B^{(s)}}{
q_{-N+s-1}q_{-N+s}},\quad \quad B^{(0)}=0
\end{equation}
and then, if we choose all integration constants zero,
\begin{equation}
B^{(s+1)}=\delta _{s,0}+\sum_{k=0}^{\infty }\left( \frac{B_{k+2}^{(s)}}{
q_{k}q_{k+1}}-\frac{B_{k}^{(s)}}{q_{k-N+s-1}q_{k-N+s}}\right) .
\label{Brecurrence}
\end{equation}
Now we prove by induction on $s$ that for $s\geq 0$
\begin{equation}
B^{(s+1)}=(-1)^{s}\sum_{k_{1}=-N}^{-1}\sum_{k_{2}=-N+1}^{k_{1}+1}\dots
\sum_{k_{j}=-N+j-1}^{k_{j-1}+1}\dots
\sum_{k_{s}=-N+s-1}^{k_{s-1}+1}Q_{k_{1}}\dots Q_{k_{s}}  \label{Bsind}
\end{equation}
where
\begin{equation}
Q_{k}=\frac{1}{q_{k}q_{k+1}}.
\end{equation}
From (\ref{Brecurrence}) we have
\begin{equation}
B^{(s+1)}=(-1)^{s+1}\beta _{1}+(-1)^{s}\beta _{2}  \label{Bs+1}
\end{equation}
where
\begin{eqnarray}
\beta _{1} &=&\sum_{k_{1}=0}^{\infty
}Q_{k_{1}}\sum_{k_{2}=-N}^{-1}\sum_{k_{3}=-N+1}^{k_{2}+1}\dots
\sum_{k_{j}=-N+j-2}^{k_{j-1}+1}\dots
\sum_{k_{s}=-N+s-2}^{k_{s-1}+1}Q_{k_{2}+k_{1}+2}\dots Q_{k_{s}+k_{1}+2}
\label{beta1} \\
\beta _{2} &=&\sum_{k_{1}=0}^{\infty
}Q_{k_{1}-N+s-1}\sum_{k_{2}=-N}^{-1}\sum_{k_{3}=-N+1}^{k_{2}+1}\dots
\sum_{k_{j}=-N+j-2}^{k_{j-1}+1}\dots
\sum_{k_{s}=-N+s-2}^{k_{s-1}+1}Q_{k_{2}+k_{1}}\dots Q_{k_{s}+k_{1}}.
\label{beta2}
\end{eqnarray}
In (\ref{beta1}) we introduce $k_{j}^{\prime }=k_{1}+k_{j}+2$ for $j=2,\dots
,s$ and renaming $k_{j}^{\prime }\rightarrow k_{j}$ we obtain
\begin{equation}
\beta _{1}=\sum_{k_{1}=0}^{\infty }\sum_{k_{2}=-N+k_{1}+2}^{k_{1}+1}\dots
\sum_{k_{j}=k_{1}-N+j}^{k_{j-1}+1}\dots
\sum_{k_{s}=k_{1}-N+s}^{k_{s-1}+1}Q_{k_{1}}Q_{k_{2}}\dots Q_{k_{s}}.
\label{beta1m}
\end{equation}
In (\ref{beta2}) we introduce $k_{1}^{\prime }=k_{1}-N+s-1$ and $
k_{j}^{\prime }=k_{1}+k_{j}$ for $j=2,\dots ,s$ and then rename $
k_{j}^{\prime }\rightarrow k_{j}$
\begin{equation*}
\beta _{2}=\sum_{k_{1}=-N+s-1}^{\infty
}\sum_{k_{2}=k_{1}-s+1}^{k_{1}-s+N}\sum_{k_{3}=k_{1}-s+2}^{k_{2}+1}\dots
\sum_{k_{j}=k_{1}+j-s-1}^{k_{j-1}+1}\dots
\sum_{k_{s}=k_{1}-1}^{k_{s-1}+1}Q_{k_{1}}\dots Q_{k_{s}}.
\end{equation*}
Exchanging the first two sums we obtain
\begin{equation*}
\beta _{2}=\sum_{k_{2}=-N}^{\infty }\sum_{k_{1}=\max
\{k_{2}+s-N,-N+s-1\}}^{k_{2}+s-1}\sum_{k_{3}=k_{1}-s+2}^{k_{2}+1}\dots
\sum_{k_{j}=k_{1}+j-s-1}^{k_{j-1}+1}\dots
\sum_{k_{s}=k_{1}-1}^{k_{s-1}+1}Q_{k_{1}}\dots Q_{k_{s}}
\end{equation*}
and with $k_{1}\leftrightarrow k_{2}$
\begin{eqnarray}
\beta _{2} &=&\sum_{k_{1}=-N}^{\infty }\sum_{k_{2}=\max
\{k_{1}+s-N,-N+s-1\}}^{k_{1}+s-1}\sum_{k_{3}=k_{2}-s+2}^{k_{1}+1}\dots
\sum_{k_{j}=k_{2}+j-s-1}^{k_{j-1}+1}\dots
\sum_{k_{s}=k_{2}-1}^{k_{s-1}+1}Q_{k_{1}}\dots Q_{k_{s}}  \notag \\
&=&\sum_{k_{1}=-N}^{-1}\sum_{k_{2}=-N+s-1}^{k_{1}+s-1}
\sum_{k_{3}=k_{2}-s+2}^{k_{1}+1}\dots
\sum_{k_{j}=k_{2}+j-s-1}^{k_{j-1}+1}\dots
\sum_{k_{s}=k_{2}-1}^{k_{s-1}+1}Q_{k_{1}}\dots Q_{k_{s}}+  \notag \\
&&+\sum_{k_{1}=0}^{\infty
}\sum_{k_{2}=k_{1}+s-N}^{k_{1}+s-1}\sum_{k_{3}=k_{2}-s+2}^{k_{1}+1}\dots
\sum_{k_{j}=k_{2}+j-s-1}^{k_{j-1}+1}\dots
\sum_{k_{s}=k_{2}-1}^{k_{s-1}+1}Q_{k_{1}}\dots Q_{k_{s}}.  \label{beta2m}
\end{eqnarray}
Let us consider first of all the second term and exchange the sums from the
left to the right. After $j-1$ inversions we have
\begin{equation*}
\sum_{k_{1}=0}^{\infty }\dots \sum_{k_{j-1}=k_{1}-N+\left( j-1\right)
}^{k_{j-2}+1}\sum_{k_{j}=k_{1}+s-N}^{k_{j-1}+s-\left( j-1\right)
}\sum_{k_{j+1}=k_{j}-s+j}^{k_{j-1}+1}\sum_{k_{j+2}=k_{j}+\left( j+2\right)
-s-1}^{k_{j+1}+1}\dots \sum_{k_{s}=k_{j}-1}^{k_{s-1}+1}Q_{k_{1}}\dots
Q_{k_{s}}
\end{equation*}
and exchanging $\sum_{k_{j}}$ and $\sum_{k_{j+1}}$ one gets
\begin{equation*}
\sum_{k_{1}=0}^{\infty }\dots \sum_{k_{j-1}=k_{1}-N+\left( j-1\right)
}^{k_{j-2}+1}\sum_{k_{j+1}=k_{1}-N+j}^{k_{j-1}+1}
\sum_{k_{j}=k_{1}+s-N}^{k_{j+1}+s-j}\sum_{k_{j+2}=k_{j}+\left( j+2\right)
-s-1}^{k_{j+1}+1}\dots \sum_{k_{s}=k_{j}-1}^{k_{s-1}+1}Q_{k_{1}}\dots
Q_{k_{s}}
\end{equation*}
so that with $k_{j}\leftrightarrow k_{j+1}$
\begin{equation*}
\sum_{k_{1}=0}^{\infty }\dots \sum_{k_{j-1}=k_{1}-N+\left( j-1\right)
}^{k_{j-2}+1}\sum_{k_{j}=k_{1}-N+j}^{k_{j-1}+1}
\sum_{k_{j+1}=k_{1}+s-N}^{k_{j}+s-j}\sum_{k_{j+2}=k_{j+1}+\left( j+2\right)
-s-1}^{k_{j}+1}\dots \sum_{k_{s}=k_{j+1}-1}^{k_{s-1}+1}Q_{k_{1}}\dots
Q_{k_{s}}
\end{equation*}
and we only have to observe that for $s=j+1$ the process ends giving for the
last sum $\sum_{k_{s}=k_{1}+s-N}^{k_{s-1}+1}$.

From (\ref{Bs+1}), (\ref{beta1m}) and (\ref{beta2m}) it then follows that
\begin{equation*}
B^{(s+1)}=\left( -1\right)
^{s}\sum_{k_{1}=-N}^{-1}\sum_{k_{2}=-N+s-1}^{k_{1}+s-1}
\sum_{k_{3}=k_{2}-s+2}^{k_{1}+1}\dots
\sum_{k_{j}=k_{2}+j-s-1}^{k_{j-1}+1}\dots
\sum_{k_{s}=k_{2}-1}^{k_{s-1}+1}Q_{k_{1}}\dots Q_{k_{s}}.
\end{equation*}
Again we exchange $\sum_{k_{2}}$ and $\sum_{k_{3}}$ getting
\begin{equation*}
B^{(s+1)}=\left( -1\right)
^{s}\sum_{k_{1}=-N}^{-1}\sum_{k_{3}=-N+1}^{k_{1}+1}
\sum_{k_{2}=-N+s-1}^{k_{3}+s-2}\sum_{k_{4}=k_{2}-s+3}^{k_{3}+1}\dots
\sum_{k_{j}=k_{2}+j-s-1}^{k_{j-1}+1}\dots
\sum_{k_{s}=k_{2}-1}^{k_{s-1}+1}Q_{k_{1}}\dots Q_{k_{s}}
\end{equation*}
and for $k_{2}$ $\leftrightarrow k_{3}$
\begin{equation*}
B^{(s+1)}=\left( -1\right)
^{s}\sum_{k_{1}=-N}^{-1}\sum_{k_{2}=-N+1}^{k_{1}+1}
\sum_{k_{3}=-N+s-1}^{k_{2}+s-2}\sum_{k_{4}=k_{3}-s+3}^{k_{2}+1}\dots
\sum_{k_{j}=k_{3}+j-s-1}^{k_{j-1}+1}\dots
\sum_{k_{s}=k_{3}-1}^{k_{s-1}+1}Q_{k_{1}}\dots Q_{k_{s}}
\end{equation*}
so that after performing all the inversions we recover (\ref{Bsind}).

\subsection*{Acknowledgments}

This work was partially supported by PRIN 2000 ``Sintesi'' and was performed
in the framework of the INTAS project 99-1782.

\end{document}